\documentclass[aps,prb,twocolumn,superscriptaddress,showpacs,floatfix]{revtex4}

\usepackage{graphicx}

\begin{document}

\title{Nanomechanics of an individual vortex in an anisotropic
type-II superconductor}

\author{E.~H.~Brandt}
\affiliation{Max-Planck-Institut f\"ur Metallforschung,
   D-70506 Stuttgart, Germany}

\author{G.~P.~Mikitik}
\affiliation{Max-Planck-Institut f\"ur Metallforschung,
   D-70506 Stuttgart, Germany}
\affiliation{B.~Verkin Institute for Low Temperature Physics
   \& Engineering, Ukrainian Academy of Sciences,
   Kharkov 61103, Ukraine}

\author{E.~Zeldov}
\affiliation{Department of Condensed Matter Physics,
Weizmann Institute of Science, Rehovot 76100, Israel}

\date{\today}

\begin{abstract}
As shown in recent experiments [Auslaender {\it et al}., Nature
Physics {\bf 5}, 35 (2009)] magnetic force microscopy permits one
not only to image but also to manipulate an individual vortex in
type-II superconductors, and this manipulation provides a new
powerful tool to study vortex dynamics and pinning. We derive
equations that describe the deformation of an individual vortex in
an anisotropic biaxial type-II superconductor under the action of
the microscope's magnetic tip. These equations take into account
the driving force generated by the tip, the elastic force caused
by the vortex deformation, and the pinning force exerted by point
defects. Using these equations, we reproduce the main features of
the experimental data obtained by Auslaender {\it et al}.
\end{abstract}


\maketitle

\section{Introduction}

In a recent paper \cite{oph} magnetic force microscopy (MFM) was
employed to image and manipulate individual vortices in a single
crystal YBa$_2$Cu$_3$O$_{6.991}$, directly measuring the
interaction of a moving vortex with the local disorder potential.
Several unexpected results were obtained in that paper. In
particular, the authors of Ref.~\onlinecite{oph} found a dramatic
enhancement of the response of a vortex to pulling when they
wiggled it transversely. In addition, they discovered enhanced
vortex pinning anisotropy in this crystal. These results
demonstrate the power of MFM to probe microscopic defects that
cause pinning and show that the described manipulations of an
individual vortex provide a new powerful tool for studying the
vortex dynamics and vortex pinning in type-II superconductors.

In this paper we derive equations that govern the vortex dynamics
under such MFM manipulations, and by solving these equations
numerically, we provide some insight into the results of
Ref.~\onlinecite{oph}.

\section{Equations for a moving vortex}

Consider a platelet-shaped biaxial anisotropic superconductor,
with its crystalline c-axis being perpendicular to the plane of
the platelet (and the a and b axes in this plane). Let there be a
vortex directed along the c-axis in the sample. We denote this
axis as the $z$ axis, and choose the $x$ and $y$ axes along the a
and b axes of the crystal. MFM employs a sharp magnetic tip placed
near the surface of the platelet. The tip magnetization exerts an
attractive force ${\bf F}$ on the vortex end. This force can shift
the top of the vortex when the tip moves. On the other hand, it is
possible to measure $\partial F_z/\partial z$ at the tip, and this
permits one to visualize the position of the top end of the
vortex. \cite{oph,Wiesendanger} Let $X$, $Y$ be the position of
the tip in the $x$-$y$ plane, while its height above the surface
of the platelet be $Z$. We shall describe the shape of the vortex
by the functions $x(z)$ and $y(z)$ with $z\le0$, the position of
the vortex end at the surface is thus $x_0\equiv x(0)$, $y_0\equiv
y(0)$. Below we shall use the following dependence of the force
${\bf F}$ on height $Z$ and on the two-dimensional vector ${\bf
R}\equiv (X-x_0, Y-y_0)$: \cite{chang,oph}
\begin{equation}\label{1}
{\bf F}=q{{\bf R}+(Z+h_0){\bf\hat z} \over (R^2+(Z+h_0)^2)^{3/2}},
\end{equation}
where the constant $h_0\approx \lambda$ ($\lambda$ is of the order
of the London penetration depth), $q=\tilde m \Phi_0/2\pi$,
$\Phi_0$ is the flux quantum, $\tilde m$ is the magnetic monopole
strength of the tip (or the magnetic moment per unit length of a 
long narrow cylinder used as tip),   
and ${\bf\hat z}$ is the unit vector along the
$z$ axis. This dependence is obtained if one considers the tip and
the end of a straight vortex  as magnetic monopoles of strengths
$\tilde m$ and $2\Phi_0 /\mu_0$. \cite{cb} The
lateral component of ${\bf F}$, ${\bf F}_{lat}$, gives the driving
force acting on the vortex. The maximum of ${\bf F}_{lat}$ with
respect to variations of $R$ is reached at $R=(Z+h_0)/ \sqrt{2}$
and is equal to \cite{c0} $F_m\approx 0.385q/(Z+h_0)^2$. In our
following numerical calculations we shall use formula (\ref{1})
even when the vortex is curved, and to describe the lateral
component ${\bf f}_{ex}^{\parallel}dz$ of the external driving
force applied to a vortex segment which has the projection $dz$ on
the $z$-axis, we shall employ the model expression
\begin{equation}\label{2}
{\bf f}_{ex}^{\parallel}=q{{\bf R} \over
(R^2+(Z+h_0)^2)^{3/2}}{\exp\left(-|z|/\lambda\right)\over
\lambda}.
\end{equation}
This expression can be justified if the change of the total
lateral force ${\bf F}_{lat}$ on the scale $\lambda$ in the
$x$-$y$ plane is relatively small (i.e., if $R\gg \lambda$).
However, in the case when the vortex shift $(x_0^2+y_0^2)^{1/2}$
caused by the tip is essentially larger than $\lambda$, this shift
is practically independent of the specific form of the
$z$-dependence of ${\bf f}_{ex}^{\parallel}$, see below. So, to
clarify the physics without additional mathematical complications,
below we shall always use the model dependences (\ref{1}) and
(\ref{2}).

As it was mentioned above, measurement of $\partial F_z/\partial
z$ is employed to visualize the vortex. Equation (\ref{1}) yields
the following expression for this derivative:
 \begin{equation}\label{3}
\left |{\partial F_z\over \partial z}\right |=
q{|2(Z+h_0)^2-R^2|\over (R^2+(Z+h_0)^2)^{5/2}}.
 \end{equation}
This derivative is maximum $|\partial F_z/\partial z|_{\rm
max}=2q/(Z+h_0)^3$ when the tip is just above the vortex, i.e.,
when $X=x_0$, and $Y=y_0$. On the other hand, the maximum lateral
force occurs when $R=(Z+h_0)/\sqrt{2}$, and hence $|\partial
F_z/\partial z|=q(2/3)^{3/2}/(Z+h_0)^3 \approx 0.27\,|\partial
F_z/\partial z|_{\rm max}$ at this $R$. In other words, the
maximum of the lateral force and the maximum of $|\partial
F_z/\partial z|$ occur at different positions of the tip and of
the vortex end.

We shall consider the vortex as an elastic string. In the case of
a biaxial superconductor the line tension of the vortex,
$\varepsilon_l(\theta,\varphi,\psi)$, and the pinning force acting
on its unit length, $f_p(\theta,\varphi,\psi)$, were calculated in
Ref.~\onlinecite{mb}. The angles $\theta$ and $\varphi$ define the
direction of the vortex, i.e., we shall describe this direction by
the unit vector
 \begin{equation} \label{4}
 (\sin\theta\cos\varphi,\sin\theta\sin\varphi,
 \cos\theta)={(x',y',1)\over \sqrt{1+x'^2+y'^2}},
 \end{equation}
while the angle $\psi$ defines the direction of the pinning force
or of the vortex distortion in the plane perpendicular to the
vortex, Fig.~1. Here the prime means $d/dz$. In the subsequent
analysis the line tension will be required only for the case
$\theta=0$ since the linear elasticity theory is valid up to
sufficiently large angles $\theta$ if the parameter $\varepsilon$
is small. \cite{eh92} Then, we have \cite{oph,mb}
\begin{eqnarray} \label{5}
\varepsilon_l(\varphi,\psi)=\varepsilon_0\varepsilon^2
\eta(\varphi+\psi)\equiv \varepsilon_l(\varphi+\psi) ,
\end{eqnarray}
where $\varepsilon\equiv \lambda_{ab}/\lambda_c$ is the parameter
of the anisotropy; $\varepsilon_0=(\Phi_0/\lambda_{ab})^2
\ln(\lambda_{ab} /\xi_{ab})/(4\pi \mu_0)$;
$\lambda_{ab}=\sqrt{\lambda_a\lambda_b}$; $\lambda_c$, $\lambda_a$
and $\lambda_b$ are the London penetration depths,
$\zeta=\lambda_a/\lambda_b$ is the parameter of the anisotropy in
the a-b plane, and
 \begin{equation}\label{6}
\eta(\varphi)=\zeta\cos^2\varphi +\zeta^{-1}\sin^2\varphi.
 \end{equation}
Since at $\theta=0$ the plane perpendicular to the vortex
coincides with the $x$-$y$ plane, the combination $\varphi+\psi$
in Eq.~(\ref{5}) is the angle defining the direction of the vortex
distortion in this plane relative to the $x$ axis. \cite{c1} As to
the pinning force, it is described by the expression \cite{mb}
 \begin{equation}\label{7}
f_p(\theta,\varphi,\psi)= f_p^c{\xi_{ab}\cos\theta\over
\xi(\theta,\varphi,\psi)},
 \end{equation}
where $f_p^c$ is the pinning force for the vortex along the c axis
in the uniaxial superconductor with the same $\lambda_{ab}$ and
$\xi_{ab}= \sqrt{\xi_a\xi_b}$. Here $\xi_a$ and $\xi_b$ are the
coherence lengths, and
\begin{eqnarray} \label{8}
\xi^2(\theta,\varphi,\psi)&=&\xi_{ab}^2\Big[\zeta (\sin\varphi
\cos\psi \cos\theta+\cos\varphi\sin\psi)^2\ \ \  \\
 &+&{1\over \zeta}(
\cos\varphi \cos\psi \cos \theta -\sin\varphi \sin\psi)^2\Big].
\nonumber
 \end{eqnarray}
In YBa$_2$Cu$_3$O$_{6.99}$ one has \cite{lambda}
$\varepsilon\approx 1/7$ (i.e., $\varepsilon^2\ll 1$) and
\cite{lambda,oph} $\zeta\approx 1.3$.

 \begin{figure}  
 \includegraphics[scale=.45]{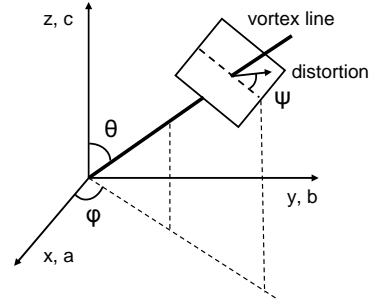}
\caption{\label{fig1} Definition of the angles $\theta$, $\varphi$
and $\psi$. The angles $\theta$ and $\varphi$ specify the
direction of the vortex shown as bold solid line. The angle $\psi$
in the plane perpendicular to the vortex defines the direction of
the pinning force; $\psi$ is measured from the line that is the
intersection of this plane with the plane containing the vortex
and the z-axis.
 } \end{figure}   

Consider a vortex segment limited by the planes $z$ and $z+dz$ and
specified by the angles $\theta$ and $\varphi$. Let us analyze the
balance of the driving, the pinning, and the elastic forces
applied to this segment. All these forces are perpendicular to it.
However, to find the two functions $x(z)$ and $y(z)$ describing
the vortex, it is convenient to carry out the analysis in the
$x$-$y$ plane, projecting all the forces onto this plane. The
projection of the elastic force acting on this segment, ${\bf
f}_{el}^{\parallel}dz$, can be described by the simple expression
${\bf f}_{el}^{\parallel}dz= (\varepsilon_{lx}x'',
\varepsilon_{ly}y'') dz$ even at sufficiently large $\theta$ since
the linear elasticity theory is valid up to the angles satisfying
$\varepsilon^2\tan^2\theta\ll 1$.\cite{eh92} Here
$\varepsilon_{lx}= \varepsilon_{l}(0)= \varepsilon_0
\varepsilon^2\zeta$ and $\varepsilon_{ly}= \varepsilon_{l}(\pi/2)=
\varepsilon_0\varepsilon^2/\zeta$ are the appropriate line
tensions at $\theta=0$, see Eqs.~(\ref{5}) and (\ref{6}). Adding
this projection of the elastic force to the external force defined
by Eq.~(\ref{2}), one obtains the projection ${\bf
f}^{\parallel}dz$ of the resultant force ${\bf f}dz$ on the
$x$-$y$ plane. Then, the {\it first} of two equations for $x(z)$
and $y(z)$ is
\begin{equation} \label{9}
  f^{\parallel}(x,y,X,Y)=f_c^{\parallel},
\end{equation}
where $f_c^{\parallel}$ is the absolute value of the projection of
the so-called critical force \cite{mb} on the $x$-$y$ plane. This
critical force is the force at which the vortex starts to move. It
is determined by the pinning force, but in the anisotropic
superconductor it can differ from the pinning force. \cite{mb}
Note that we write one equation (\ref{9}) which connects the
absolute values of ${\bf f}^{\parallel}$ and ${\bf
f}_c^{\parallel}$ rather than two equations for the $x$ and $y$
components of these forces. This is due to the fact that the
direction of the pinning force (and hence of the critical force)
is not known in advance and is dictated by the direction of ${\bf
f}^{\parallel}$.

The critical force $f_c^{\parallel}$ is determined by the
following formulas: Let the direction of the force ${\bf f}$ be
specified by the angle $\psi$ in the plane perpendicular to the
vortex. This angle can be expressed in terms of the component
$f_x^{\parallel}$ and $f_y^{\parallel}$ of the force ${\bf
f}^{\parallel}$ as follows:
\begin{equation}\label{10}
  \tan\psi
  ={\cos\theta(f_y^{\parallel}-f_x^{\parallel}\tan\varphi)
  \over f_x^{\parallel}+f_y^{\parallel}\tan\varphi}.
\end{equation}
The pinning force $f_p$ in the direction $\psi$ is given by
Eqs.~(\ref{7}) and (\ref{8}), while the critical force $f_c$ in
this direction $\psi$ is determined by \cite{mb}
 \begin{eqnarray}\label{11}
\tan(\psi-\psi_1)= {f_p'(\psi_1)\over f_p(\psi_1))}, \\
f_c(\psi)=\sqrt{[f_p(\psi_1)]^2 + [f_p'(\psi_1)]^2 }, \label{12}
 \end{eqnarray}
where the prime means $d/d\psi_1$, and the angle $\psi_1$ in the
plane perpendicular to the vortex defines the direction along
which the vortex starts to move when the force acting along $\psi$
exceeds $f_c$. The fact that $\psi_1$ in general differs from
$\psi$ is due to the anisotropy of the pinning. On determining
$\psi_1$ from Eq.~(\ref{11}), one then finds $f_c(\psi)$ from
formula (\ref{12}). The explicit form of Eqs.~(\ref{11}) and
(\ref{12}) for the case of the pinning force described by formulas
(\ref{7}) and (\ref{8}) is presented in Appendix \ref{A}. Finally,
when the critical force $f_c(\psi)$ is found, its projection
$f_c^{\parallel}$ on the $x$-$y$ plane is determined by the
formula
\begin{equation}\label{13}
  f_c^{\parallel}=f_c(\psi){(\cos^2\theta \cos^2\psi+
  \sin^2\psi)^{1/2} \over \cos\theta},
\end{equation}
that follows from geometrical considerations.

Equation (\ref{9}) is a differential equation since it contains
the derivatives $x''(z)$ and $y''(z)$ originating from the elastic
force. As in Ref.~\onlinecite{oph}, we shall consider only
sufficiently thick superconducting crystals in which the vortex as
a whole does not shift, and only its upper part ($0\ge z \ge z_0$)
adjoining the $x$-$y$ surface moves, while the lower part ($z <
z_0$) is pinned. The boundary point $z_0$ of this upper part is
determined by
\begin{equation}\label{14}
  x(z_0)=y(z_0)=0.
\end{equation}
Then, the boundary conditions to Eq.~(\ref{9}) are
\begin{eqnarray}\label{15}
  x'(z_0)&=&y'(z_0)\!=0,  \\
  x'(0)&=&y'(0)\,=0. \label{16}
\end{eqnarray}
If these conditions were not fulfilled, the derivatives $x'$ and
$y'$ would be discontinuous at the points $z=z_0$ and $z=0$, and
the elastic force $(\varepsilon_{lx}x'', \varepsilon_{ly}y'')$
would be singular there. \cite{c2} In the most interesting case
when $x_0^2+y_0^2\gg \lambda^2$ (and hence $|z_0|\gg \lambda$),
one can put $\lambda \to 0$. In this limiting case the driving
force ${\bf F}$ is applied to the vortex only at its surface point
($x_0$,$y_0$). Then, in equation (\ref{9}) the force ${\bf
f}_{ex}^{\parallel}$ can be omitted, ${\bf f}^{\parallel}$
coincides with ${\bf f}_{el}^{\parallel}$, and the driving force
${\bf F}$ only modifies the boundary condition (\ref{16}). Now the
integration of the forces over the thickness of the surface layer
gives
\begin{equation}\label{17}
  x'(0)={F_x \over \varepsilon_{lx}},\ \ \
  y'(0)={F_y \over \varepsilon_{ly}}.
\end{equation}
This result shows that at small $\lambda$ the vortex dynamics is
practically independent of the distribution of the driving force
${\bf F}$ over the surface layer of thickness $\lambda$.

Equation (\ref{9}) alone is not sufficient to find the two
functions $x(z)$ and $y(z)$. We now derive a {\it second} equation
for these functions. When the position of the tip changes, the
vortex begins to move in the direction $\psi_1$ in the plane
perpendicular to the vortex. This movement of the vortex in the
perpendicular plane corresponds to its shift at an angle $\tilde
\psi_1$ (measured from the $x$ axis) in the $x$-$y$ plane. A
geometrical consideration shows that this angle $\tilde \psi_1$ is
determined by
\begin{equation}\label{18}
\tan \tilde \psi_1={\tan\varphi +\cos\theta \tan\psi_1
   \over 1-\cos\theta \tan\varphi \tan\psi_1}.
\end{equation}
Thus, changes of the functions $x(z)$ and $y(z)$ in time are
connected by the relation
 \begin{equation}\label{19}
{dy\over dt}=\tan\tilde \psi_1 {dx \over dt}.
 \end{equation}
This is a second equation for the functions $x(z)$ and $y(z)$.
Since the time $t$ can be expressed in terms of the known
functions $X(t)$, $Y(t)$ that describe the shift of the tip,
Eq.~(\ref{19}) and its solution (i.e., the shape of the vortex at
some moment $t_0$) depend on the {\it trajectory} $Y(X)$ of the
tip in the $x$-$y$ plane at previous times ($t<t_0$) rather than
on a specific form of the temporal dependences $X(t)$ and $Y(t)$.
This situation is reminiscent of the case that occurs in the
theory of the critical states of type-II superconductors when the
external magnetic field ${\bf H}_a$ applied to a superconducting
sample changes in a complex way.\cite{crst,crst1} In this case the
critical states are different for different histories ${\bf
H}_a(t)$ with the same final value of ${\bf H}_a$.

Equations (\ref{1})-(\ref{19}) describe the vortex dynamics in
thick superconducting crystals when the tip moves in its $x$-$y$
plane. We solve these equations in the next section.

\section{Results}

The equations of the previous section show that if the
driving-force density $f_{ex}^{\parallel}$ at the surface of the
superconductor, $z=0$, is lower than a certain threshold
$f_c^{\parallel}(\alpha)$ where $\alpha$ is the angle of ${\bf
f}_{ex}^{\parallel}$ relative to the $x$ axis, the vortex remains
pinned, i.e., $x(z)=y(z)=0$. In particular, if the driving force
acts along the $x$ or $y$ direction, we obtain the following
thresholds: $f_p^c\sqrt{\zeta}$ and $f_p^c/\sqrt{\zeta}$,
respectively, which coincide with the appropriate pinning forces.
Here we have used the formulas of Appendix A and the fact that
$\delta=1/\zeta^2>1/2$ at the experimental value \cite{lambda,oph}
of $\zeta=1.3$. Equivalently, these threshold conditions can be
rewritten in terms of the total forces, $F_x \le F_{px}\equiv
f_p^c \lambda \sqrt \zeta$  and $F_y\le F_{py}\equiv f_p^c
\lambda/\sqrt \zeta$. If the driving force exceeds the threshold
values only a little, i.e., if $F_x-F_{px}\ll F_x$, or
$F_y-F_{py}\ll F_y$, we find from the equations that $x_0$ or
$y_0$ begins to deviate gradually from zero,
 \begin{equation}\label{20}
x_0\approx{2\lambda (F_x-F_{px})^3\over \varepsilon_{lx} F_x^2}, \
\ \ y_0\approx{2\lambda(F_y-F_{py})^3\over \varepsilon_{ly}F_y^2}.
 \end{equation}
With further increase of the driving force, at $F_x\gg F_{px}$ or
$F_y\gg F_{py}$ but at the same time under the conditions $F_x \ll
\varepsilon_{lx}=\zeta \varepsilon^2\varepsilon_0$ or $F_y \ll
\varepsilon_{ly}=\varepsilon^2\varepsilon_0/\zeta$, we arrive at
 \begin{equation}\label{21}
x_0\approx{F_x(F_x-2F_{px})\over 2\zeta^{3/2} f_p^c
\,\varepsilon^2\varepsilon_0}, \ \ \
y_0\approx{\zeta^{3/2}F_y(F_y-2F_{py})\over 2f_p^c
\,\varepsilon^2\varepsilon_0}.
 \end{equation}
The additional conditions $F_x \ll \varepsilon_{lx}$,  $F_y \ll
\varepsilon_{ly}$ mean that the characteristic tilt angle $\theta$
of the vortex is small [see Eqs.~(\ref{17}) in which $x'(0)$,
$y'(0)$ are just equal to $\tan\theta$]. This smallness of
$\theta$ was assumed in Ref.~\onlinecite{oph} in analyzing the
vortex dynamics, and formulas (\ref{21}) coincide with those
obtained in that paper. However, $F_{px}/\varepsilon_{lx}$,
$F_{py}/\varepsilon_{ly}$ are not necessarily small in an
experiment. In this case formulas (\ref{21}), strictly speaking,
have no region of applicability. Moreover, boundary conditions
(\ref{17}) show that the characteristic tilt angle $\theta$ is not
small at typical experimental values of $F_{x,y}\sim 5-20$ pN even
when $\varepsilon_{lxy}\sim 10$ pN. So we do not assume in this
paper that $\tan\theta\ll 1$. The equations of the previous
section have been derived only under a weaker condition
$\varepsilon^2\tan^2\theta \ll 1$. But when $\theta \sim 1$, the
critical force $f_c$ differs from $f_p$ even for symmetry
directions.\cite{mb} For example, when the tip moves along $x$ and
thus the vortex also bends along this direction, formula
(\ref{A5}) of Appendix A gives
\begin{eqnarray}\label{22}
  f_c^{\parallel}(\theta)\!\!\!&=&\!\!f_p^c\sqrt \zeta,\ \ \
  \tan^2\theta \le {2\over \zeta^2}-1; \\
f_c^{\parallel}(\theta)\!\!\!&=&\!\!{2f_p^c\over
\zeta^{3/2}}\cos\theta \sqrt{\zeta^2-\cos^2\theta},\ \ \
  \tan^2\theta \ge {2\over \zeta^2}-1, \nonumber
\end{eqnarray}
while for the tip moving along the $y$ axis, one has
\begin{eqnarray}\label{23}
f_c^{\parallel}(\theta)\!\!\!&=&\!\!\!{f_p^c\over \sqrt \zeta},\ \
\  \tan^2\theta \le 2\zeta^2-1; \\
f_c^{\parallel}(\theta)\!\!\!&=&\!\!\!2\zeta^{1/2}\!f_p^c\cos\theta
\sqrt{1\!-\!\zeta^2\cos^2\theta},\ \ \
  \tan^2\theta \ge 2\zeta^2-1.~~ \nonumber
\end{eqnarray}
In other words, even at moderate $\theta$ the critical force
begins to depend on this angle, and the formulas for $x_0$ and
$y_0$ become more complicated than Eqs.~(\ref{21}) in which
$f_c^{\parallel}$ was assumed to be constant and to coincide with
the appropriate pinning force, $f_c^{\parallel}(\theta)=f_p^c\sqrt
\zeta$ at $\varphi=0$ and $f_c^{\parallel}(\theta)=f_p^c/\sqrt
\zeta$ at $\varphi=\pi/2$. Such a dependence of
$f_c^{\parallel}(\theta)$ in general causes that the ratio
$y_0/x_0$ at large driving forces differs from the value
$\zeta^3\approx 2.2$ that follows from formulas (\ref{21}). This
may lead to an imitation of the enhanced pinning anisotropy
observed by Auslaender {\it et al}.,\cite{oph} see below.

 \begin{figure}  
 \includegraphics[scale=.45]{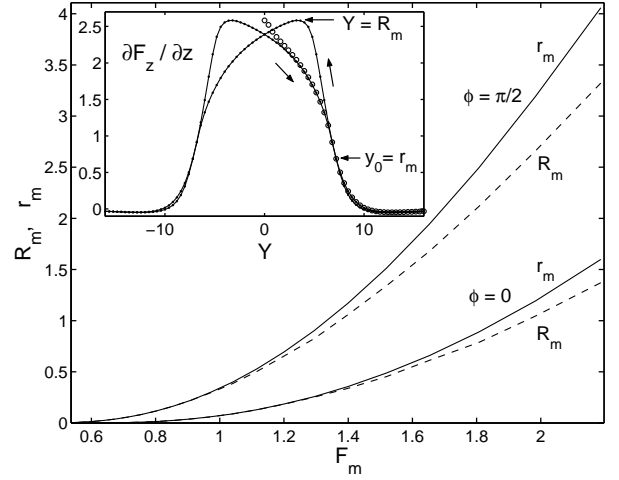}
\caption{\label{fig2} Dependence of the maximum shift of the
vortex end, $r_m\equiv {\rm max}(x_0^2+y_0^2)^{1/2}$, on the
driving force $F_{mx}$ or $F_{my}$ when the tip moves along the
$x$ axis ($\varphi=0$) or the $y$ axis ($\varphi=\pi/2$). The
forces are measured in units of the line tension
$\varepsilon_{lxy}\equiv \varepsilon^2\varepsilon_0$, the lengths
in units of $\lambda$, $\zeta=1.3$, and $P\equiv
f_p^c\lambda/\varepsilon_{lxy}=0.5$. The dashed lines show the
appropriate $X$ and $Y$ positions of the tip, $R_m\equiv
(X^2+Y^2)^{1/2}$, at which the derivative $\partial F_z/\partial
z$ reaches its maximum on the returning path. As an example, the
$Y$-dependence of this derivative at $\phi=\pi/2$ and
$F_{my}/\varepsilon_{lxy}\approx 2.2$ is presented in the inset.
The circles in the inset mark the virgin curve, and the arrows
indicate the direction of the tip motion.
 } \end{figure}   

During its motion the vortex lags behind the moving tip until the
maximum lateral force is reached at $r_m={\rm max}
(x_0^2+y_0^2)^{1/2}$. At small driving force the vortex will
remain at this $r_m$, whereas at large driving forces the vortex
will, in fact, partially recede after the tip has moved away.
Experimentally the final location of the vortex is evaluated on
the returning path of the tip by monitoring the tip location
$R_m\equiv (X^2+Y^2)^{1/2}$ at which $\partial F_z/\partial z$ is
maximum when the tip is above the vortex (or closest to it). In
Fig.~\ref{fig2} we show the maximum shift of the vortex end,
$r_m$, in the forward direction and $R_m$ on the returning path of
the tip vs. the driving force when the tip moves either along the
$x$ axis or along the $y$ axis. In these cases the vortex shifts
along these symmetric directions, too. Figure~\ref{fig2} shows
that at low driving forces the experimentally determined $R_m$
accurately reproduces the maximum shift of the vortex $r_m$
whereas at higher forces $R_m$ slightly underestimates the actual
$r_m$.

\begin{figure}  
 \includegraphics[scale=.45]{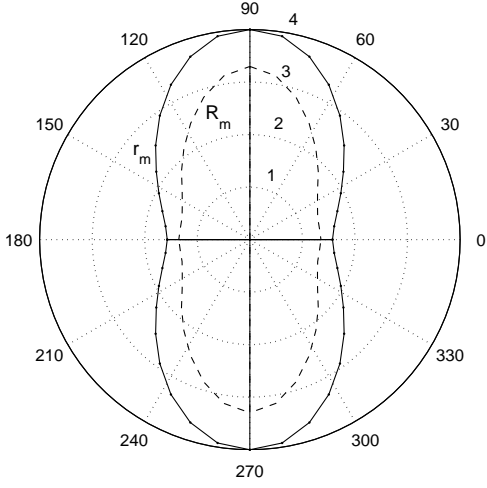}
\caption{\label{fig3} Polar plot of the maximum shift $r_m$ of the
vortex end $r=(x_0^2 +y_0^2)^{1/2}$ versus the angle $\varphi$ of
the tip motion (solid line). The tip moves along a straight line
in the $X$-$Y$ plane with a sufficiently large amplitude and at
the height $Z$ that leads to $F_m/\varepsilon_{lxy}\approx 2.2$.
The dashed line shows the shift $R_m=(X^2+Y^2)^{1/2}$ of the tip
when the derivative $\partial F_z/\partial z$ reaches its maximum.
Length unit is $\lambda$, $P=0.5$, $\zeta=1.3$.
 } \end{figure}

\begin{figure}  
 \includegraphics[scale=.45]{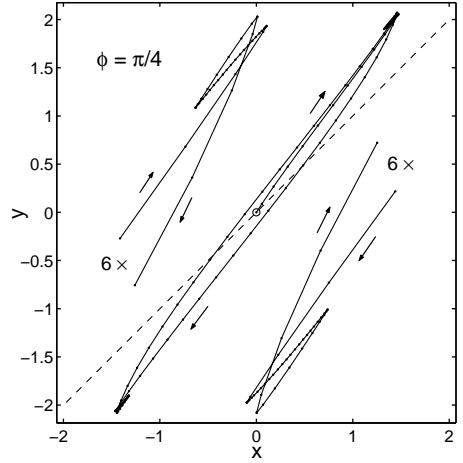}
\caption{\label{fig4} Path of the vortex end $x(t)$, $y(t)$ when
the tip oscillates along the diagonal $X(t)=Y(t)$ (dashed line)
with a large amplitude; same data as in Fig.~\ref{fig3}. Both the
tip and the vortex start at $x=y=0$. Length unit is $\lambda$,
$P=0.5$. The vortex path cycles a narrow hysteresis loop as
indicated by the arrows. Due to the in-plane anisotropy $\zeta =
1.3$, this loop is tilted away from the tip-path ($\varphi=\pi/4$)
towards the $y$ axis. Also depicted is the six times enlarged and
shifted path near the first and the second turns. The dots on the
curves are at equidistant times.
 } \end{figure}

\begin{figure}  
 \includegraphics[scale=.45]{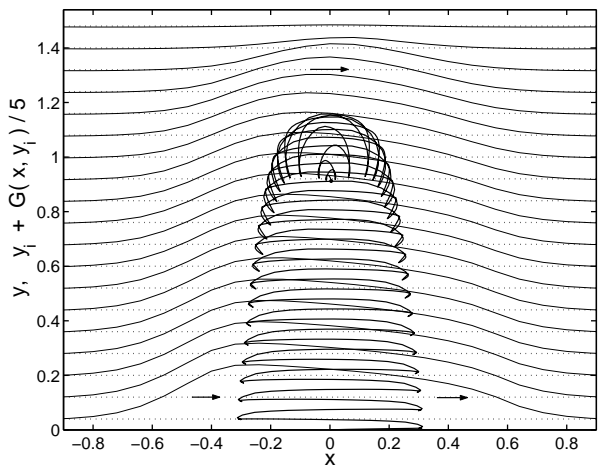}
\caption{\label{fig5} The zigzag path $x_0(t)$, $y_0(t)$ of the
vortex end (bold lines in the center) when the tip oscillates with
large amplitude $a=1.6\,\mu$m along $x$ and at the same time moves
slowly up along $y$, with $\dot Y /| \dot X| =1/80$. Tip and
vortex start at $x=y=0$. Length unit is $\mu$m, $\zeta=1.3$, the
other parameters are listed in Eqs.~(\ref{24}). The aspect ratio
of this path is max($y_0$)/max($x_0$) $\approx 3.7$. The almost
horizontal dotted lines at equidistant $y = y_i$ show the tip path
when it moves from the left to the right (see arrows) and serve as
zero lines for the force derivative $g(x,y_i) =\partial
F_z/\partial z$ plotted versus $x$ as $y_i + 0.2 \cdot G(x,y_i)$
(solid lines) with $G=g/{\rm max}(|g|)$. Note that these curves
are asymmetric due to the unidirectional tip motion shown here.
The return path yields similar curves, obtained from the depicted
curves by the reflection $x\to -x$.
 } \end{figure}

\begin{figure}  
 \includegraphics[scale=.45]{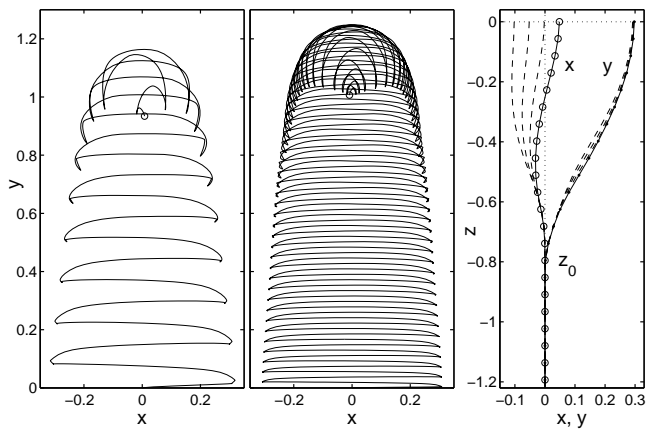}
\caption{\label{fig6} The zigzag path $x_0(t)$, $y_0(t)$ of the
vortex end as in Fig.~\ref{fig5} but at $\dot Y /| \dot X| =1/40$
(left plot, max($y_0$)/max($x_0$) $\approx 3.6$) and at $\dot Y /|
\dot X| =1/160$ (middle plot, max($y_0$)/max($x_0$)=4.1). The
right plot shows the vortex shape expressed as  $x(z)$ (solid line
with circles) and $y(z)$ (solid line with dots) at the moment when
$x_0=0.05$, $y_0=0.3$ in the left plot. The dashed lines show
these functions at three previous time steps.
 } \end{figure}

\begin{figure}[t]  
 \includegraphics[scale=.45]{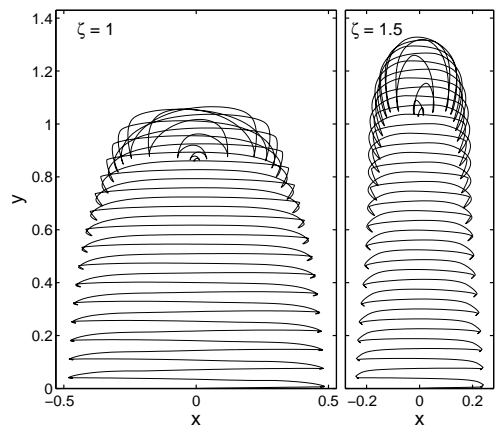}
\caption{\label{fig7} The zigzag path $x_0(t)$, $y_0(t)$ of the
vortex end as in Fig.~\ref{fig5} but for $\zeta=1$ (left plot)
and  $\zeta =1.5$ (right plot). The aspect ratio
max($y_0$)/max($x_0$) is approximately 2.2 for
$\zeta=1$ and 5.5 for $\zeta=1.5$.
 } \end{figure}

\begin{figure}[t]  
 \includegraphics[scale=.45]{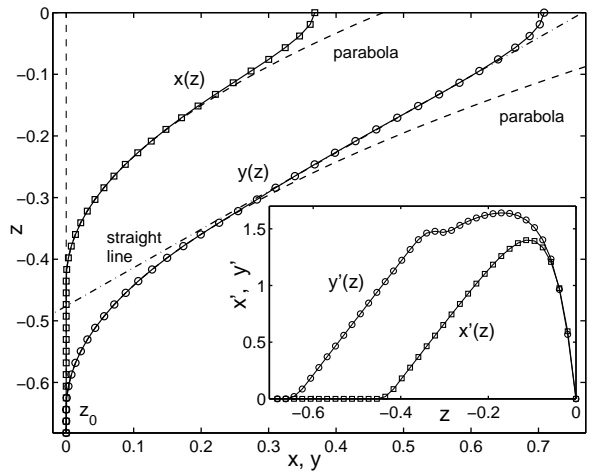}
\caption{\label{fig7a} The vortex shape during oscillations of the
magnetic tip above a superconductor that is isotropic in the a-b
plane ($\zeta=1$). Similar case as the left plot of
Fig.~\ref{fig7}, but to clarify the situation, we take
$\lambda_{ab}=0.05\,\mu$m and $P=0.25$ here. Shown are the maximum
vortex displacement $x(z)$ at the first excursion of the tip
(i.e., at $Y=0$) and the maximum displacement $y(z)$ at the moment
when $x_0=0$ while $y_0$ reaches its maximum value after many tip
oscillations. The dash-dotted straight line reveals that the curve
$y(z)$ has a long zero-curvature segment, see also $x'(z)$ and
$y'(z)$ shown in the inset (the small hump seen in the flat part
of $y'(z)$ oscillates in time). At small $x$ and $y$ both $x(z)$
and $y(z)$ are parabolas with curvature $f_p^c /
\varepsilon_{lxy}$ (dashed lines).
 } \end{figure}

\begin{figure}[tbh]  
 \includegraphics[scale=.50]{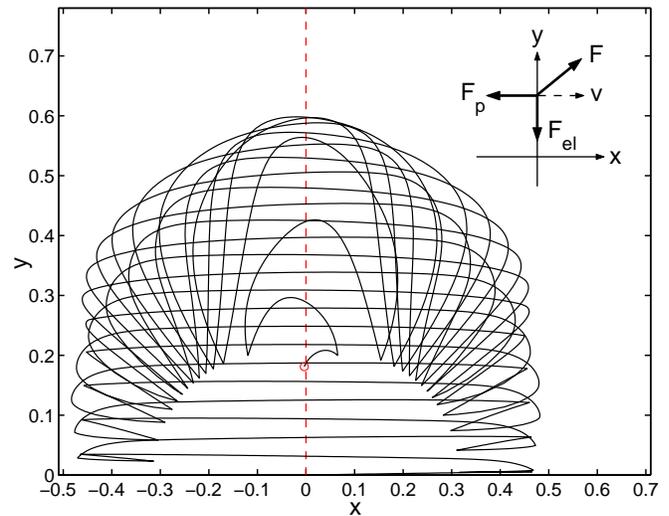}
\caption{\label{fig7b} Path of the vortex end and force balance
for a simplified 2D model, see text. The tip moves as in
Fig.~\ref{fig7}, the left plot. Here $Z+h_0=0.44\ \mu$m, $q=9.9\
\mu$m$^2\cdot$pN (which gives $F_m\approx 20$ pN); $F_p=F_m/4$;
$\zeta=1$; $k_x=k=32$ pN/$\mu$m; $x$ and $y$ are measured in
$\mu$m. The aspect ratio $r\equiv {\rm max}(y_0)/ {\rm
max}(y_0)\approx 1.24$. The force balance is shown for the point
$(x_0,y_0)=(0,{\rm max}(y_0))$.
 } \end{figure}

\begin{figure}[tbh]  
 \includegraphics[scale=.50]{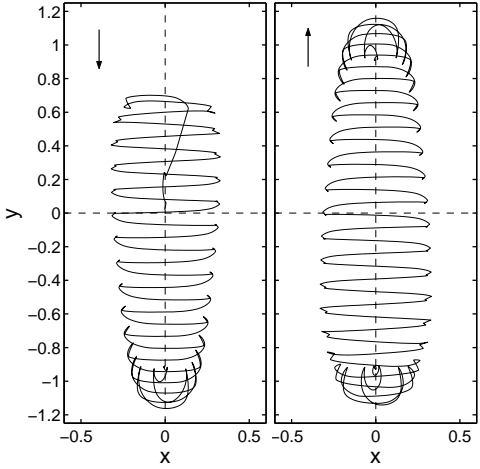}
\caption{\label{fig8}  Path of the vortex end when the tip
oscillates along $x$ with large amplitude and moves down from
large positive $Y \ge 2$ to large negative $Y \le -2$ (left plot)
and then moves up again to large positive $Y \ge 2$ (right plot).
The straight vortex waits at $x=y=0$. When the tip approaches from
above, the vortex end suddenly jumps to the tip and starts to
oscillate with large amplitude, following the tip downwards. After
some time the vortex end comes to a halt as in Figs.~\ref{fig5},
\ref{fig6}, and \ref{fig7}. When the oscillating tip approaches
again from below, the vortex end starts to oscillate with slowly
increasing amplitude along a path that looks similar to the path
on which the vortex end came to a halt. The vortex paths shown at
the lower left and at the upper right are nearly identical. The
parameters are as in the left plot of Fig.~\ref{fig6}.
 } \end{figure}

\begin{figure}[tbh]  
 \includegraphics[scale=.409]{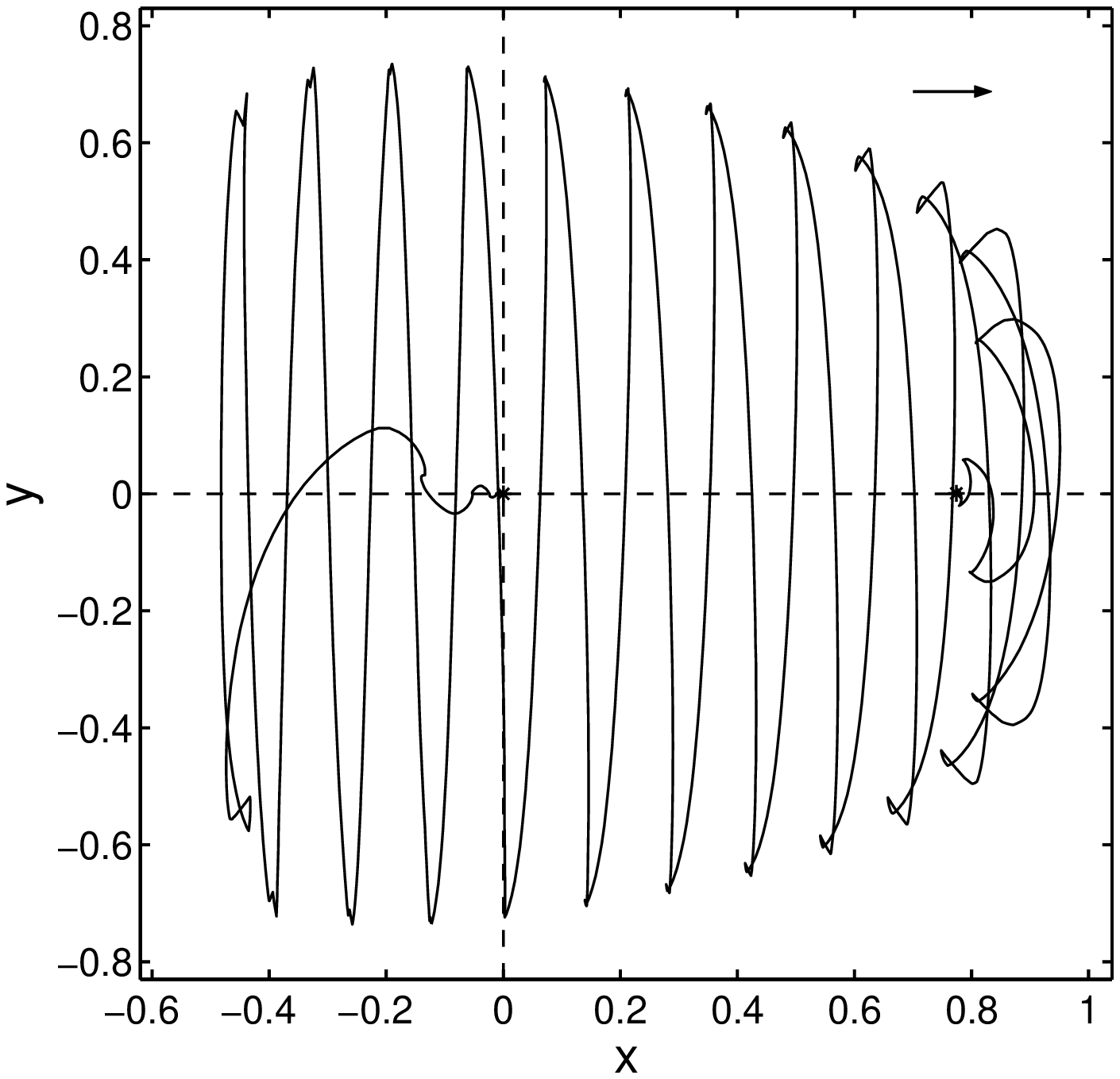}
\caption{\label{fig9} Path of the vortex end for parameters as in
the left plots of Figs.~\ref{fig6} and ~\ref{fig8}, but now the
tip oscillates along $y$ and approaches the vortex end (that waits
at $x=y=0$) along $x$ from far left ($X \le -2$), moving further
until the vortex end comes to a halt.
 } \end{figure}

\begin{figure}[tbh]  
 \includegraphics[scale=.45]{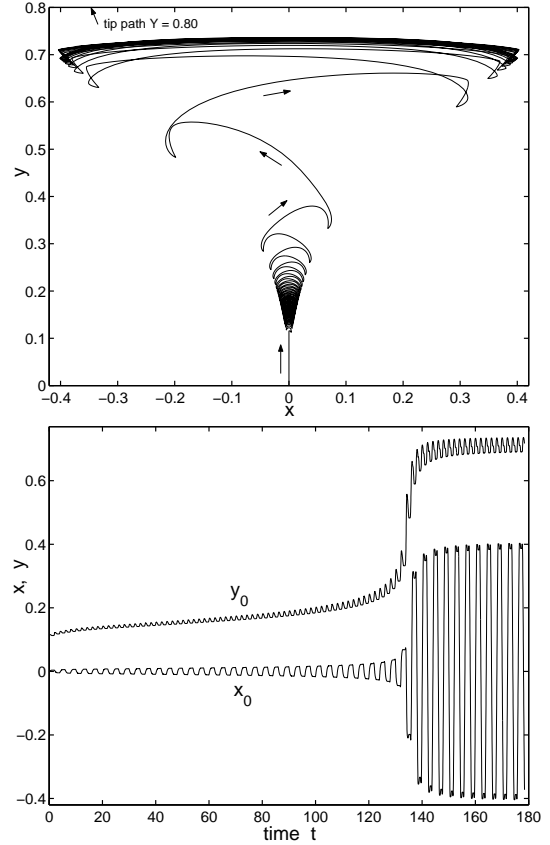}
\caption{\label{fig12} Attraction of the vortex end to the
oscillating tip. The magnetic tip oscillates with amplitude
$a=1.4$ along the straight line  $Y=0.8$ parallel to the $x$ axis,
starting from $X=0$ at time $t=0$. When the tip approaches the
starting point from large positive $Y$, the vortex end shifts to
$y_0 \approx 0.11$, attracted by the tip. At $t>0$ the vortex end
oscillates along $x$ with small, slightly increasing amplitude,
moving slowly to higher $y$ values. When $y_0 \approx 0.3$ is
reached the vortex end jumps in a few big leaps to its maximum
$y_0 \approx 0.73$. After that it oscillates on a stationary
curve. The lower plot shows the temporal dependences of $x_0$ and
$y_0$. All lengths in $\mu$m, the unit of $t$ is a quarter of the
tip period, the parameters are as in Fig.~\ref{fig8}, but for
simplicity we take $\zeta=1$ here.
 } \end{figure}

In the construction of Fig.~\ref{fig2}, as well as
Figs.~\ref{fig3} and \ref{fig4}, we put $\zeta=1.3$ and measure
forces in units of the line tension $\varepsilon_{lxy} \equiv
(\varepsilon_{lx} \varepsilon_{ly})^{1/2} = \varepsilon^2
\varepsilon_0$, and lengths in units of $\lambda$ (hence the force
densities $f_p$, $f_c$, $f_{el}$, and $f_{ex}^{\parallel}$ are in
units of $\varepsilon_{lxy}/ \lambda$). Then, taking into account
the model dependence (\ref{2}) for the driving force density
$f_{ex}^{\parallel}$, one finds that equations (\ref{9}) and
(\ref{19}) for $x(z)$ and $y(z)$, as well as the boundary
condition (\ref{17}), become independent of the absolute values of
$\varepsilon_0$, $\varepsilon$, and $\lambda$. They depend only on
the dimensionless forces $F_{x,y}/\varepsilon^2\varepsilon_0$ and
the dimensionless parameter $P\equiv f_p^c\lambda / \varepsilon^2
\varepsilon_0$. Thus, in a certain sense Fig.~\ref{fig2} is
universal. But in this scaling procedure one has to bear in mind
that if one changes the parameter $\lambda$ keeping a fixed value
of $F_{x,y}/\varepsilon^2\varepsilon_0$, this leads to a change of
the tip position $X$, $Y$ which is not scaled with $\lambda$, see
Eq.~(\ref{1}). However, if one is interested only in the tip
position when it is just above the vortex, the scaling still holds
in this case. On the other hand, when the relative positions of
the tip and of the vortex are essential (e.g., in the construction
of Figs.~\ref{fig5} -\ref{fig12}), we use the following set of
input parameters:
\begin{eqnarray}\label{24}
\lambda\!\!\!&=&\!\!0.2\,\mu{\rm m},\ \ \varepsilon_{lxy}\!\!
\equiv \!(\varepsilon_{lx} \varepsilon_{ly})^{1/2}\!=9\,{\rm
pN}, \nonumber \\
P\!\!&=&\!\!{f_p^c\lambda\over \varepsilon_{lxy}}\!=\!0.5,\ \
{q\over \varepsilon_{lxy}}\!\!=\!1.1\,\mu{\rm m}^2,\ \
Z\!+\!h_0\!\!=\!0.44\,\mu{\rm m}.~~~~
\end{eqnarray}

The data of Fig.~\ref{fig2} are similar to the data of Fig.~3b in
Ref.~\onlinecite{oph}. Moreover, a semiquantitative agreement of
these data can be obtained if one takes $\lambda$ of the order of
several tenths of a micron and $\varepsilon_{lxy}\sim 10$ pN.
However, this value of the line tension $\varepsilon_{lxy}$ is
$10-20$ times larger than the theoretical estimate of this
quantity, $\varepsilon_{lxy} = \varepsilon^2
(\Phi_0/\lambda_{ab})^2 \ln(\lambda_{ab} /\xi_{ab})/(4\pi \mu_0)$,
at $\varepsilon=1/7$, $\lambda_{ab}=0.2\,\mu$m,
$\ln(\lambda_{ab}/\xi_{ab})= 4$. Thus, apart from an enhanced
anisotropy of pinning discovered by Auslaender {\it et
al}.,\cite{oph} their experimental data in fact means that either
the vortex has an enhanced line tension, or the model dependences
(\ref{1}) and (\ref{2}) for the driving force are oversimplified
under the conditions of the experiment and lead to an essential
overestimation of this force.

In Fig.~\ref{fig3} that is similar to Fig.~4c of
Ref.~\onlinecite{oph}, we show the  dependence of the maximum
shift of the vortex end, $r_m$, on the angle $\varphi$ at which
the tip moves along a straight line in the $x$-$y$ plane at a
certain height $Z$ above the surface of the superconductor. This
height determines the maximum driving force $F_m$ applied to the
vortex, and in Fig.~\ref{fig3} this height is chosen so that
$F_m/\varepsilon_{lxy}\approx 2.2$. For comparison, we again show
the positions of the tip, $R_m\equiv (X^2+Y^2)^{1/2}$, at which
the derivative $\partial F_z/\partial z$ reaches its maximum. The
anisotropy of the vortex shift, $r_m(\varphi=\pi/2)/
r_m(\varphi=0)\approx 2.5$, seen in the figure approximately
coincides with the ratio $R_m(\varphi=\pi/2)/R_m(\varphi=0)$, and
at $\zeta=1.3$ this anisotropy is lower than the appropriate
experimental value $\sim 3.5$.\cite{oph} This experimental value
can be fitted if one takes $\zeta=1.43$. Thus, although this
$\zeta=1.43$ obtained with taking into account the
$\theta$-dependence of $f_c^{\parallel}$ is less than $\zeta=1.6$
derived in the simplified analysis \cite{oph}, our approach still
cannot completely describe the enhanced anisotropy of pinning
within the framework of collective pinning theory by point
defects. Auslaender {\it et al}. \cite{oph} suggested that this
enhanced anisotropy is due to a clustering of the point defects.

Interestingly, when the tip moves along a straight line different
from the $x$ and $y$ axes, the trajectory of the vortex end
performs a ``hysteresis loop'' with its axis deviating from the
direction of tip motion, Fig.~\ref{fig4}. Also depicted in
Fig.~\ref{fig4} is the six times enlarged path near the first and
the second turns, showing that the vortex end reaches maximum
elongation, then it recedes when the tip moves away, and when the
tip returns, the vortex end approaches the tip and reaches maximum
elongation a second time. These results clearly demonstrate that
the vortex in general moves in a direction different from the
direction of the tip motion, and that the vortex position depends
on the trajectory of the tip at previous times.

In Ref.~\onlinecite{oph} the derivative $(\partial F_z/\partial
z)$ was measured when the tip oscillates with a large amplitude
along some line and at the same time it is slowly shifted in the
perpendicular direction. In this case an enhanced shift of the
vortex along the slow scan direction was discovered, see Figs.~1
and 2 in Ref.~\onlinecite{oph}. We have investigated this
situation theoretically. In Fig.~\ref{fig5} the zigzag path
$x_0(t)$, $y_0(t)$ of the vortex end is presented when the tip
oscillates with a large amplitude along $x$ and at the same time
moves slowly up along $y$. We also show the $X$ profiles of
$(\partial F_z/\partial z)$ at various fixed values of $Y$. Note
that these profiles are asymmetric and are different for tip
motion from left to right and from right to left. The data of
Fig.~\ref{fig5} qualitatively reproduces the experimental data.
\cite{oph} Interestingly, this figure also clearly shows how the
elastic force drags the vortex back towards the origin when the
tip goes far away from the vortex.

In Fig.~\ref{fig6} we compare the vortex paths for various ratios
of the scan rates along $x$ and $y$. It is seen that the results
are different for different rates although we do not take into
account the effect of vortex creep here. This difference in the
vortex paths is due to the above-mentioned dependence of the
vortex position on the trajectory of the tip at previous times. In
this figure we also present the vortex-shape functions $x(z)$ and
$y(z)$ at some moment of time. These functions show that during
the zigzag motion the vortex is bent and twisted into a
complicated shape. The lower part $z \le z_0$ of the vortex is
rigidly pinned (has exactly $x=y=0$) and at the surface $z = 0$
the vortex ends perpendicularly. We find that for the tip motion
of Fig.~\ref{fig6}, at $z > z_0$ the component $y(z)$ increases
with $z$ monotonically, while $x(z)$ after several scan periods
exhibits strongly damped oscillations.

In Fig.~\ref{fig7} we analyze the dependence of the zigzag vortex
motion on the anisotropy parameter $\zeta$. It is clear from the
figure that the shift of the vortex end in the slow scan direction
and the aspect ratio max($y_0$)/max($x_0$) increase \cite{c3} with
increasing $\zeta$. But importantly, even in the case of isotropic
pinning in the $x$-$y$ plane, i.e., at $\zeta=1$, this aspect
ratio remains considerably larger than unity. From a qualitative
point of view, this enhanced tilt of the vortex along $y$ is
caused by the fact that during the zigzag motion the vortex
predominately moves in the $x$ direction, the pinning force is
also directed mainly along $x$, and hence this force opposes only
the vortex tilt in the $x$ direction. These considerations are
supported by the data of Fig.~\ref{fig7a} in which for the case of
a small $\lambda$ we show $x(z)$, the maximum displacement of the
vortex when the tip moves only along the $x$ axis (i.e., during
the first oscillation of the tip in the left plot of
Fig.~\ref{fig7}), and $y(x)$ at the moment when $y_0$ reaches its
maximum value after many oscillations of the tip. Since at small
$\lambda$ the driving force concentrates near the surface of the
superconductor, in the bulk of the sample the elastic force
associated with the curvature of the vortex has to be balanced
mainly by the pinning force. Then, the {\it long straight segment}
of the line $y(z)$ shown in Fig.~\ref{fig7a} means that the $y$
component of the pinning force is practically absent in this
segment.

Some insight into the origin of the vortex-motion anisotropy seen
in Fig.~\ref{fig7} can be also obtained from a simplified
two-dimensional (2D) model. In this model we disregard the
dynamics of the entire vortex and consider only the vortex end as
a point ($x_0$,$y_0$) elastically connected to the origin of the
$x$-$y$ plane, ${\bf F}_{el}=-(k_x x_0,k_y y_0)$, where $k_x$ and
$k_y=k_x/\zeta^2$ are some spring constants modelling the
elasticity of the vortex. In this simplified approach the problem
of the vortex motion becomes two-dimensional, and instead of the
force densities we deal with the elastic force ${\bf F}_{el}$, the
total pinning force ${\bf F}_p$, and the driving force ${\bf
F}_{lat}=(F_x,F_y)$ determined by Eq.~(\ref{1}). The balance of
these three forces and the vortex-end motion can be still
described by the equations of Sec.~II if one puts $\theta=\varphi
= 0$ and replaces the force densities by the total forces in the
equations. Interestingly, in this simplified 2D approach one can
qualitatively reproduce the main results which have been obtained
above accounting for the real 3D shape of the vortex. In
particular, in Fig.~\ref{fig7b} we show the zigzag path of the
vortex end in the case $\zeta=1$ (isotropic elasticity and pinning
in the $x$-$y$ plane). In the construction of this figure we use
the same parameters for the tip as in Fig.~\ref{fig7} (i.e., we
have $F_m\approx 20$ pN). Besides this, we take $F_p=F_m/4\approx
5$ pN. This relation also corresponds to the case of
Fig.~\ref{fig7} if one implies that $F_p$ for the 2D model is
equal to $f_p^c\lambda$. Such choice of $F_p$ is dictated by a
comparison of the conditions $F_m\ge F_p$ and $F_m\ge
f_p^c\lambda$ for a vortex to start to move in the simplified 2D
model and in the three-dimensional theory. The spring constant in
Fig.~\ref{fig7b} is chosen such that ${\rm max}(x_0)$ is the same
as in the left plot of Fig.~\ref{fig7}. The vortex trajectory
presented in Fig.~\ref{fig7b} reveals the anisotropy of the vortex
motion in the $y$ and $x$ directions with the aspect ratio
$r\equiv {\rm max}(y_0)/{\rm max}(x_0) \approx 1.24$. This
anisotropy can be understood from the following simple
considerations: The maximum displacement of the vortex end along
$x$ is found from
 \begin{equation}\label{25}
{\rm max}(x_0)={F_m-F_p\over k},
 \end{equation}
where $F_m$ is the maximum value of the driving force and $k\equiv
k_x$. The $y_0$ reaches its maximum when $x_0\approx 0$, the
vortex-end velocity $v$ is practically parallel to $x$, and thus
the pinning force is along this axis, too, Fig.~\ref{fig7b}. The
driving force at this moment is maximum, $F=F_m$, and is directed
at an angle $\alpha$ with respect to the $x$ axis, while the
elastic force acts towards the origin. Then, the force balance for
the $x$ and $y$ components gives
 \begin{equation}\label{26}
F_m\cos\alpha=F_p, \ \ \  F_m\sin\alpha=k\, {\rm max}(y_0),
 \end{equation}
and hence ${\rm max}(y_0)=F_m\sin\alpha/k  =\sqrt{F_m^2 -
F_p^2}/k$. The aspect ratio is therefore
 \begin{equation}\label{27}
r={{\rm max}(y_0)\over {\rm max}(x_0)}=\sqrt{F_m+F_p\over F_m-F_p}
>1 ,
 \end{equation}
and it is independent of $k$. If $F_m \to F_p$ the ratio $r$
diverges, but in this case the vortex displacements are small and
become less than the vortex radius which is of the order of
$\lambda_{ab}$ for MFM. For $F_m$ and $F_p$ of Fig.~\ref{fig7b}
formula (\ref{27}) yields the aspect ratio $r=\sqrt{5/3}\approx
1.29$, which is indeed close to that found in this figure.

In Fig.~\ref{fig8} we analyze one more effect that was observed
experimentally, see Figs.~4c and 4d in Ref.~\onlinecite{oph}. At
the initial time $t=0$, the straight vortex is at  $x=y=0$. The
tip oscillates along $x$ with a large amplitude and slowly
approaches the vortex from large positive $y$. At a certain time
the end of the vortex abruptly jumps to the tip and then begins to
oscillate with a large amplitude. This effect of a sharp onset of
the signal is qualitatively reproduced by our Fig.~\ref{fig8}. A
close look to Fig.~\ref{fig8} shows that the large jump of the
vortex end is composed of several jumps of width increasing nearly
exponentially in time. These multiple jumps are even better seen
in Fig.~\ref{fig9} that shows how the vortex end moves when the
tip oscillates along $y$ and slowly moves along $x$ starting far
away from the waiting vortex. As compared to the corresponding
Figs.~\ref{fig5}, \ref{fig6}, and \ref{fig8} which are described
by the same parameters and have a vortex-path aspect ratio
max($y_0$)/max($x_0$)$\approx 4$, in Fig.~\ref{fig9} the aspect
ratio max($x_0$)/max($y_0$) $\approx 1.3$ is smaller than even
that for the isotropic case ($\approx 2.2$) since the pinning
anisotropy now impedes \cite{c3} the vortex motion in the $x$
direction.

\begin{figure}[tbh]  
 \includegraphics[scale=.45]{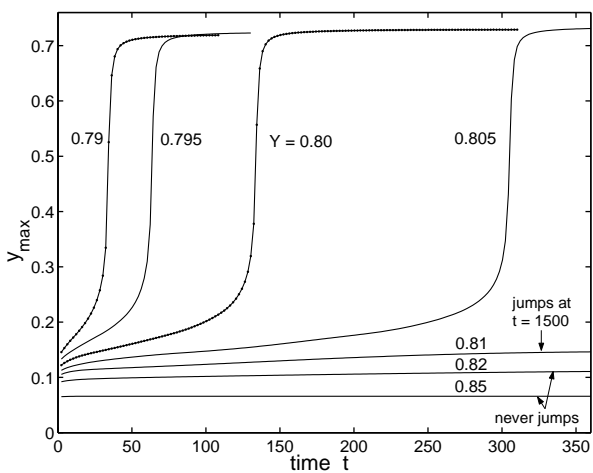}
\caption{\label{fig13} Attraction of the vortex end $(x_0,y_0)$ to
the tip oscillating along $x$ at constant $Y$ as in
Fig.~\ref{fig12} but for various distances $Y=0.79, \dots 0.85$.
Plotted is the maximum value $y_{\rm max}$ of $y_0$ in each half
oscillation vs. time $t$. For $0.79 \le Y \le 0.81$ this $y_{\rm
max}$ is slowly increasing and then suddenly jumps to a saturation
value $\approx 0.73$ within about five half oscillations. At
$Y=0.81$ this steep jump occurs only at $t=1500$ (after $375$
oscillations). For $Y \ge 0.815$,  $y_{\rm max}$ saturates
exponentially in $t$ to a small value $\le 0.114$, and thus there
is no jump. All parameters and units are the same as in
Fig.~\ref{fig12}.
 } \end{figure}   

In Fig.~\ref{fig12} we reproduce one more experiment described in
Ref.~\onlinecite{oph} (in the supplementary material). In this
experiment the tip oscillates along a straight line at $t>0$ and
does not shift in the perpendicular direction. The vortex that
waits at some distance from the line of the tip oscillations at
$t\le 0$, at $t>0$ begins to move towards the tip.
Figure~\ref{fig12} shows this attraction process for the isotropic
case $\zeta=1$ and for the tip-oscillations line parallel to the
$x$ axis. The initial shift $y_0(0)$ of the vortex end along $y$
occurs at $t \le 0$ when the tip approaches its starting point
from large positive $Y$. This shift occurs if the driving force at
$t=0$ exceeds the appropriate pinning force $F_p=f_p^c\lambda$. If
the initial distance of the vortex from the tip-oscillations line
is so large that the driving force is less than this pinning
force, the vortex end remains pinned and does not move towards the
tip. A more restrictive necessary condition for the vortex motion
towards the tip is that the vortex can oscillate along $x$. This
condition yields
 \begin{equation}\label{28}
(Y-y_0(0))^2 \le - {(Z+h_0)^2\over 2}+\sqrt {{(Z+h_0)^4\over 4}+
{q^2\over 9 F_p^2}}.
 \end{equation}
From numerical calculations we find, see Fig.~\ref{fig13}, that
there exists a distinct upper threshold for the distance
$(Y-y_0(0))$ between the vortex end and the tip-oscillation line
at which the attraction process can occur, and this threshold is
close to that given by Eq.~(\ref{28}).\cite{c4} If this threshold
is indeed determined by the pinning forces and the dependence of
the driving force $F$ on $X-x_0$ and $Y-y_0$ is known, this effect
may allow sensitive measurements of these pinning forces acting on
an individual vortex in type-II superconductors.

When the tip oscillates, it generates currents whose orientation
changes in time near the vortex, and the vortex motion towards the
tip in Fig.~\ref{fig12} as well as the enhanced vortex response in
the slow scan direction in Figs.~\ref{fig5}-\ref{fig7} are
reminiscent of the so-called longitudinal vortex shaking effect.
\cite{lshake} In this effect, in essence, a small ac current is
superimposed perpendicularly to a dc critical current that flows
in a sample. This leads not only to a periodic tilt of vortices
but also to their {\it unidirectional} drift along the direction
of the ac current and causes a dc electric field along the dc
current. In the considered case of the oscillating tip the
currents flow only near the surface of the superconductor, and
only the upper, depinned part of the vortex ``drifts''.

\section{Conclusions}

We derive equations that describe the deformation of an individual
vortex in anisotropic type-II superconductors under the influence
of the moving tip of a magnetic force microscope. These equations
take into account the driving force generated by the tip, the
elastic force caused by the vortex deformation, and the pinning
force exerted by point defects. These equations are valid even at
large deformations of the vortex, and they properly allow for the
biaxial anisotropy of the superconductor. From these equations, we
reproduce the main features of the experimental data obtained
recently.\cite{oph} In particular, we explain the enhanced
response of the vortex to pulling in the slow scan direction as
compared to its response in the direction of the fast zigzag scan.
We demonstrate that the vortex position at time $t$ depends on the
trajectory of the tip at previous times, and it is this property
that eventually leads to the enhanced vortex response in the slow
scan direction. We also point out that the enhanced anisotropy of
pinning in the $a$-$b$ plane that was observed in
Ref.~\onlinecite{oph} is partly caused by the fact that the
critical force at which the vortex starts to move depends on the
angle $\theta$ of the vortex tilt and in general does not coincide
with the pinning force.

We note a still unresolved problem. In order to obtain
quantitative agreement of our calculations with the experimental
data, we have to take a larger value of the vortex line tension
than the value following from the theoretical estimate. The small
line tension $\sim \varepsilon^2 \varepsilon_0$ of a vortex in an
anisotropic bulk superconductor results from the almost complete
cancellation of the increase of the length of a tilted vortex and
the decrease of its energy per unit length, $e_l(\theta)\approx
\varepsilon_0\cos\theta$, with increasing tilt angle $\theta$.
\cite{eh92} The existence of the surface at $z=0$ and of the tip
changes the energy $e_l(\theta)$ in the surface layer of depth
$\lambda$ and, consequently, the almost complete cancellation does
not occur there. The line tension of a vortex segment near the
surface may thus be noticeably larger than the tension in the
bulk. The discrepancy also may be due to the too simple
expressions for the lateral driving force, see Eqs.~(\ref{1}) and
(\ref{2}). Since the penetration depth $\lambda$ of the driving
force should be of the order of $\lambda_{ab}$, this $\lambda$ is
comparable with the experimental values of $Z+h_0$. In this
situation the correct driving force acting on a {\it curved}
vortex at small distances $R\sim Z+h_0$ from the tip, is likely to
be given by formulas more complicated than Eqs.~(\ref{1}) and
(\ref{2}). But Eq.~(\ref{1}) was, in fact, used in the experiment
\cite{oph} for the extraction of the lateral driving force, which
might lead to some overestimation of this force. Thus, a more
detailed theoretical investigation of the driving force and the
nonlocal line tension near the surface is needed.

One more problem that should be studied both theoretically and
experimentally is the vortex-motion randomness that is
superimposed on the regular vortex motion considered here. This
randomness is clearly seen in the experimental data of Auslaender
{\it et al}.\cite{oph} It is quite possible that apart from point
defects and the weak collective pinning associated with them, in
the sample there may be strong pinning centers, e.g., the clusters
of point defects discussed in Ref.~\onlinecite{oph}, that lead to
the observed randomness.

\acknowledgments

We thank Ophir Auslaender for discussions and for providing data.
This work was supported by the German Israeli Research Grant
Agreement (GIF) No G-901-232.7/2005. EZ acknowledges the support of
EU-FP7-ERC-AdG and of US-Israel Binational Science Foundation (BSF).

\appendix

\section{Formulas for the critical force}\label{A}

Taking into account formulas (\ref{7}) and (\ref{8}), we obtain
the following explicit form of Eqs.~(\ref{11}) and (\ref{12}):
 \begin{eqnarray}\label{A1}
\tan\psi&=&{N(\psi_1)\over D(\psi_1)}, \\
f_c(\psi)&=&f_p(\psi_1)\Big[1+{[\xi'(\theta,\varphi,\psi_1)]^2\over
\xi^2(\theta,\varphi,\psi_1)}\Big]^{1/2}, \label{A2}
 \end{eqnarray}
where $f_p(\psi_1)=f_p(\theta,\varphi,\psi_1)$ is given by
Eq.~(\ref{7}), the prime means $d/d\psi_1$,
 \begin{eqnarray}\label{A3}
N\!\!&=&\!\!\eta\tan^3\psi_1+1.5\Delta \zeta \tan^2\psi_1
\cos\theta
\sin2\varphi  \nonumber \\
&+&\tan\psi_1 (2\eta_1\cos^2\theta-\eta)-0.5\Delta \zeta
\cos\theta \sin2\varphi, \nonumber \\
D\!\!&=&\!\!\eta_1\cos^2\theta\!+\! 1.5\Delta \zeta\tan\psi_1
\cos\theta \sin2\varphi \\
&+&\!\!\tan^2\!\psi_1 (2\eta-\eta_1\cos^2\theta)\!-0.5 \Delta
\zeta \tan^3\!\psi_1 \cos\theta \sin2\varphi, \nonumber \\
{|\xi'|\over \xi}\!\!\!&=&\!\!\!{\Delta \zeta\cos\theta
\sin2\varphi\cos 2\psi_1 +\sin 2\psi_1(\eta -\cos^2\theta\eta_1)
\over \Delta \zeta \cos\theta \sin2\varphi\sin 2\psi_1
\!\!+\!2\cos^2\psi_1\cos^2\theta\eta_1\!\!+\!2\sin^2\psi_1 \eta},
 \nonumber
 \end{eqnarray}
and $\Delta \zeta\equiv \zeta\!-\!\zeta^{-1}$, $\eta\equiv
\eta(\varphi)$, $\eta_1\equiv \eta(\varphi+\pi/2)$. Equation
(\ref{A1}) permits one to find the auxiliary angle $\psi_1$
in terms of $\psi$, and
then one can calculate $f_c$ from Eq.~(\ref{A2}).

Let us define the parameter $\delta$ by the formula
 \begin{equation}\label{A4}
2\delta\!\!\equiv\!\!\left({\eta \over
\cos\theta}\!+\!\eta_1\cos\theta\!-\! \sqrt{\left({\eta \over
\cos\theta}\!+\!\eta_1\cos\theta \right)^2\!-\!1}\
\right)^2\!\!\!.~~~~~~~
 \end{equation}
For example, if $\varphi=0$, one has $\delta=
\cos^2\theta/\zeta^2$, while if $\varphi=\pi/2$, we obtain
$\delta= {\rm min}[\cos^2\theta\zeta^2,
(\cos^2\theta\zeta^2)^{-1}]$. If the parameter $\delta$ is larger
than $1/2$, there is a one-to-one correspondence between $\psi$
and $\psi_1$. At $\delta <1/2$ the situation changes.\cite{mb} In
this case spurious branches of $\psi_1(\psi)$ appear. The physical
branch corresponds to a minimum value of $f_c$. For example, in
the case $\delta \le 1/2$ one finds the following expression for
the critical force at $\psi=0$ and $\varphi=0$ or $\varphi=\pi/2$:
 \begin{equation}\label{A5}
f_c(\psi=0;\varphi=0,\pi/2)=2f_{p0}\sqrt{\delta(1-\delta)},
 \end{equation}
where $f_{p0}=f_p^c\sqrt\zeta$ at $\varphi=0$ and $f_{p0}=f_p^c/
\sqrt \zeta$ at $\varphi=\pi/2$. On the other hand, at $\delta \ge
1/2$ one has $f_c(\psi=0;\varphi=0,\pi/2)=f_{p0}$ instead of
formula (\ref{A5}).


{}


\begin{thebibliography}{}

\bibitem{oph} O.M. Auslaender, L. Luan, E.W.J. Straver,
J.E. Hoffman, N.C. Koshnick, E. Zeldov, D.A. Bonn, R. Liang, W.N.
Hardy, and K.A. Moler, Nature Physics {\bf 5}, 35 (2009).

\bibitem{Wiesendanger} U. H. Pi, Z. G. Khim, D.H. Kim, A. Schwarz,
M. Liebmann, and R. Wiesendanger, \apl {\bf 85}, 5307 (2004).

\bibitem{chang} A.M. Chang, H.D. Hallen, L. Harriott, H.F. Hess,
H.L. Kao, J. Kwo, R.E. Miller, R. Wolfe, J. van der Ziel, and
T.Y. Chang, Appl.\ Phys.\ Lett.\ {\bf 61}, 1974 (1992).

\bibitem{cb} G.~Carneiro and E.~H.~Brandt, \prb{\bf 61}, 6370 (2000).

\bibitem{c0} Since the maximum $F_{zm}$ of the force component $F_z$
occurs at $R=0$, Eq.~(\ref{1}) gives $F_m\approx 0.385F_{zm}$.

\bibitem{mb} G.~P.~Mikitik and E.~H.~Brandt, \prb{\bf 79},
 020506(R) (2009).

\bibitem{eh92} E.~H.~Brandt, \prl{\bf 69}, 1105 (1992).

\bibitem{c1} The use of formula (\ref{5}) means that we neglect
effects of the sample surface, $z=0$, and of the tip on the line
tension of the vortex segment in the surface layer of thickness
$\lambda$.

\bibitem{lambda} T. Pereg-Barnea, P.J. Turner, R. Harris, G.K.
Mullins, J.S. Bobowski, M. Raudsepp, R. Liang, D.A. Bonn, and
W.N. Hardy, \prb{\bf 69}, 184513 (2004).

\bibitem{c2} At $z=0$ it is necessary to take into account the
image of the vortex, and Eqs.~(\ref{16}) mean the absence of a
sharp break in $x'(0)$ and $y'(0)$. Note that we allow for the
effect of the sample surface on the vortex dynamics only in these
boundary conditions.

\bibitem{crst} G.~P.~Mikitik and E.~H.~Brandt, \prb{\bf 71},
012510 (2005).

\bibitem{crst1} E.~H.~Brandt and G.~P.~Mikitik, \prb{\bf 76},
064526 (2007).

\bibitem{c3} If the tip oscillates along $y$ and slowly moves
in the $x$ direction, the shift of the vortex end in the slow scan
direction $x$ and the aspect ratio ${\rm max}(x_0)/{\rm max}(y_0)$
{\it decrease} with increasing  anisotropy $\zeta$.

\bibitem{c4} Equation (\ref{28}) gives $Y-y_0(0) \le 0.8$
for the parameters of Fig.~\ref{fig12}, while the data of
Fig.~\ref{fig13} reveal the threshold $Y-y_0(0)\approx 0.73$.

\bibitem{lshake} G.~P.~Mikitik and E.~H.~Brandt, \prb{\bf 67},
104511 (2003).

\end{thebibliography}
\end{document}